# PERFORMANCE OF TURBO CODED OFDM UNDER THE PRESENCE OF VARIOUS NOISE TYPES

SPYRIDON K. CHRONOPOULOS*, VASILIS CHRISTOFILAKIS*,
GIORGOS TATSIS, PANOS KOSTARAKIS

*Physics Department, University of Ioannina, Panepistimioupolis, Ioannina, 45110, Greece*

* Corresponding authors. E-mail address: schrono@cc.uoi.gr, vachrist@uoi.gr

**Abstract.** A telecommunication system uses carriers in order to transmit information through a cable or wirelessly. If each time only one carrier is transmitted, then the system's signal will not be immune to frequency selective fading. If frequency selective fading includes the working frequency of the system, then the wireless link will not be established. Orthogonal Frequency Division Multiplexing (OFDM) is the primary solution for coping with inter-signal interference and frequency-selective fading. Many carriers can be produced by splitting a fast information stream to slower data series. Different orthogonal frequencies carry slower data series. System's performance can be further enhanced with the utilization of Turbo Codes. Turbo codes make the system more immune to noise effects with excellent BER results. This paper presents the thorough analysis of a Turbo Coded OFDM scheme using a PCCC technique in the presence of a channel which includes AWGN, Phase noise (PN), Rayleigh fading, Rician fading and Doppler shift.

# 1. Introduction

OFDM scheme has been established as the natural choice for a variety of existing and future telecommunication standards mainly due to its robustness and spectral efficiency [1-2]. An OFDM scheme can produce various orthogonal modulated sinusoidal signals which correspond to different frequencies. These different frequencies carry a portion of the original information stream that has been split amongst them. The previous separation is conducted with the help of serial to parallel conversion before IFFT. If the line of sight (LOS) doesn't exist and frequency selective fading occurs then severe problems will not be caused to the system because only a small percentage of frequencies (information subcarriers) will have been lost. Additionally, the guard band (cyclic prefix – CP) helps avoiding intersymbol interference (ISI). Also, zero padding (ZP) can be added to the system for assuring that IFFT input is a power of two. Moreover, the ZP and CP boost system performance with a drawback of reducing the rate of the data transmission [3-5].

Turbo coding is another significant innovation amongst others (such as UWB and tunable antennas) [6-8] as it exhibits excellent performance against severe noise effects. Turbo coding techniques include two kinds of schemes which are based on the way the included convolutional encoders are placed in relation to each other. These schemes are CCCs that stand for Concatenated Convolutional Codes. When two codes are concatenated in parallel then PCCC codes are produced and if they are concatenated serially then SCCC are produced. In our previous work, a PCCC scheme along with our new iterative decoding section [9] was merged with a designed OFDM platform [10] in order to form an innovative system of Turbo Coded OFDM (TC-OFDM) [11]. The new system which can produce a large number of subcarriers has already been evaluated in the presence of an AWGN channel, and moreover the fluctuation of its power output was studied in terms of PAPR (Peak to Average Power Ratio) [12-13]. The motivation of this research was to evaluate the performance of the TC-OFDM system for more realistic scenarios under severe types of noise. Each scenario included not only one type of noise but even the combination of various types of severe noises for the purpose of identifying the exact boundaries of our system's performance. The types of noise included (apart from AWGN) various effects such as phase noise, Rayleigh fading, multipath Ricean fading channel and Doppler shift. ITU multipath intensity profile of a Pedestrian channel was included in multipath Ricean fading channel simulations along with Rayleigh fading processes [14]. Phase noise (PN) was inserted in the noise scenario by accepting the fact that it could be originated from amplifier's non-linear characteristics and RF components' imperfections [15]. As for the Doppler shift, this is the relevant physical quantity to the time variance of a mobile radio channel and it is used in the simulations with the form of maximum Doppler spread [16].

This paper is split into six sections. The second section presents the Turbo Coded OFDM (TC-OFDM) system while the third section presents the innovative Turbo Codes which were used. The fourth section shows the various noise types which were utilized in our simulated platform. Finally, the fifth section presents the simulation settings along with the performance results of the TC-OFDM in the presence of diverse noise conditions. The sixth is relevant to the conclusions and future scopes of this research.

## 2. Turbo Coded OFDM System

The simulation platform consisted of three modules. The first module was the transmitter. In the transmitter module, a binary generator produced random data while afterwards turbo coding applied to these data. The turbo encoded information passed through several blocks for reaching the zero padding process where zeros were used in the end and the beginning of the signal's frame in order this signal to pass through IFFT. This block ensures the orthogonality of the produced subcarriers with a best possible efficient total bandwidth. Another advantage is the lack of need for N oscillators (for N transmitted carriers) [17-18].

It should be mentioned that an alteration is conducted in the position of ZP (Fig. 1). For example, if the OFDM subcarriers range from 0 to 2N, then zeros are placed in the middle of the subcarriers (which are the region N). This transformation is intended for placing zeros in the middle of the subcarriers as shown in Fig 1. The idea of the transformation originates from the protocol 802.11a. According to the Nyquist sampling theorem, if the sampling frequency equals to $f_S$ then the maximum detected frequency without having folding effects (aliasing) [19] is at most equal to $f_s/2$. The zeros were placed in the middle of subcarriers (Fig. 1) for the reason of allocating the subcarriers frequencies up to $f_s/2$. Obviously there will be negative frequencies up to -$f_s/2$ which in practice do not exist. The CP (cyclic prefix) is appended to the signal after the IFFT in order to provide annotation against intersymbol interference. Finally, unbuffered signal (parallel to serial conversion) is conducted and the signal is driven through the channel (which is the second stage of the simulation with various noises). The channel includes AWGN, additional phase noise (PN), Rician fading with Rayleigh fading processes, and Doppler shift.

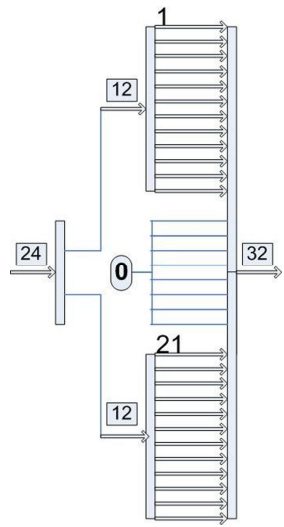

**Fig. 1. Frame transformation before IFFT**

The receiver is the third module of the simulation platform which includes a buffering process for producing a parallel information stream originated from a serial noisy channel. The cyclic prefix is removed, the signal passes through FFT and then zero padding is removed. The processes of unbuffering and buffering are conducted for creating the proper signal frames. Finally, the signal is demodulated and by using unipolar to bipolar conversion it is driven to the decoder's section where at its output the produced data will be compared to the original information coming from the binary generator.

It is worth mentioning that there was no need for synchronizing the system of transmitter-receiver because the total delay was inserted as a parameter in the simulations. Also, a graphical user interface was created in order to provide the capability of easily changing various system's values and especially carriers' number.

## 3. Efficiency of Turbo Codes

Turbo codes exhibit outstanding performance, literally changing a system's behavior in a unique way. Various schemes exist and they are characterized by the way of code's concatenation. If the code's concatenation is performed in a parallel way then PCCC codes [20-21] are produced and if it happens serially then SCCC codes are constructed. Our previous studies included the development of new PCCC codes [9] which they were compared with existing and high-performance typical systems of PCCC and SCCC codes. Specifically, the new Turbo codes in the encoder's part consisted of three convolutional encoders joined in parallel. One encoder was accepting the data from the binary generator while the other two were accepting its interleaved versions. After concatenation, a function of transposing, reshaping and frame conversion was conducted (Fig.2).

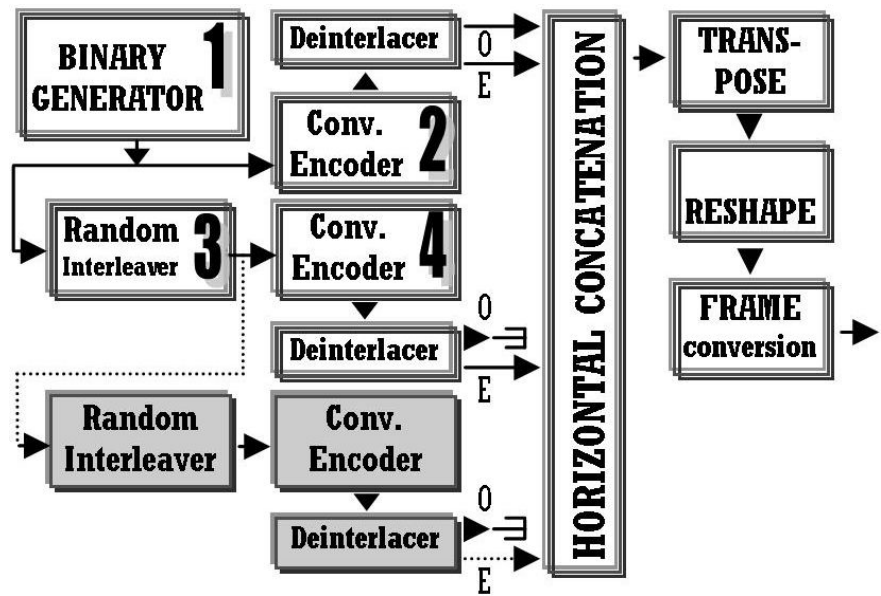

Fig. 2. Parallel and serial Turbo encoder (SCCC is the sequence 1234)

A decoder's part [22] is consisted of various puncturing blocks in order to acquire the proper stream and in turn to drive it to the appropriate interlacer (Fig.3). Every interlacer (three in total) is joined with the corresponding input in iterative section. Our designed section consists of three APP decoders. Notably the third in line decoder is joined by the feedback loop with the first one and with a hard decision block, where is conducted the final choice of bit selection (whether is "1" or "0") as appearing in Fig. 4. Then the decoder output is guided to the evaluation section (acquisition of BER performance).

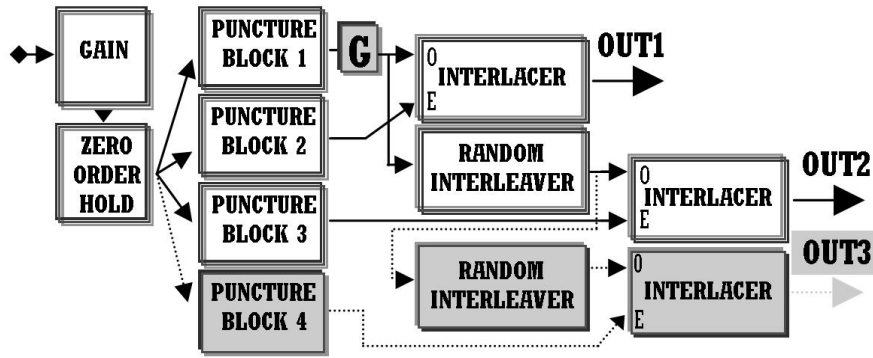

Fig. 3. Preliminary stage of PCCC decoding (gray blocks are needed for the new PCCC scheme)

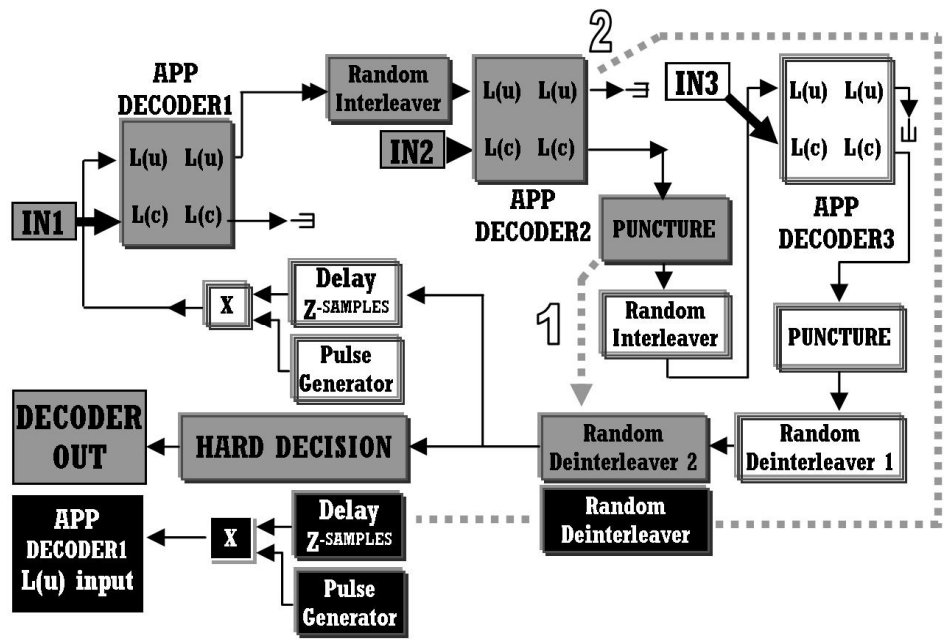

Fig. 4. Final stage of PCCC decoding (Typical design includes grey and black blocks including dashed routes 1 and 2, while the new decoder is presented by all white and grey blocks without the inclusion of dashed routes).

# 4. Simulated environment and parameters

This section describes a channel with various noise types which were utilized in the simulated scenario. Different noise types included AWGN, Phase noise (PN), Rayleigh fading, Rician fading and Doppler shift.

The term "channel" declares the medium which surrounds everything between transmitter's and receiver's antennas. When a signal is transmitted and it travels towards the receiver's antenna it changes due to the existing obstacles (e.g. buildings) and due to the route of broadcast. Nevertheless, the condition of the received signal can be totally retrieved by applying specific techniques. This is accomplished as long as the medium between transmitter-receiver has been accurately modeled and its behavior is known to the receiver. Consequently the "channel model" is the term which characterizes the medium [23].

The Additive White Gaussian Noise (AWGN) channel is considered in simulation scenarios as white noise. Also, the term "Additive" means that the addition of noise is conducted based on Equation (1). The "White" noise has a power spectral density (W/Hz) which is constant. Also, "Gaussian" refers to the distribution of noise amplitudes. [24-26].

$$y(t) = x(t) + n(t) \quad 0 \leq t \leq T \tag{1}$$

where y(t) is the resultant signal, x(t) corresponds to the transmitted signal, n(t) refers to the added AWGN samples with a power spectral density of $N_0/2$ (W/Hz) and T equals to the duration of the transmitted symbol.

A type of impairment which is called phase noise appears as random signal frequency fluctuations. If an oscillator is selected as an example for the comprehension of the phase noise, then Equation (2) should be used. The terms $V_o$ and f represent the amplitude and frequency of the produced signal respectively. The instantaneous output can be described by Equation (3) where q(t) corresponds to the phase fluctuations of the signal. The term q(t) consists of two types of noises which are parasitic signals (spurious signals) and the phase noise. The noise level calculation is measured in $dB_c/Hz$ and corresponds to the displacement (offset) of the carrier's frequency. If $P_s$ is the power of the carrier and $P_{ssb}$ is the power of the frequency offset for a bandwidth equal to 1 Hz, then the phase noise is described by the Equation (4) [27]. A(t) stands for amplitude fluctuations.

$$V(t) = V_O \sin(2\pi f t) \tag{2}$$

$$V(t) = V_O [1 + A(t)] \sin[(2\pi f t) + q(t)] \tag{3}$$

$$S_C(f) = \frac{P_S}{P_{SSB}} \Leftrightarrow S_C(f)(dB) = 10\log\frac{P_S}{P_{SSB}} \tag{4}$$

The spectral characteristics of the received signal y(f) may be calculated if the characteristics of the channel are already known, i.e. the channel's impulse response H(f) and the spectral characteristics of the transmitted signal x(f). If the transmitted signal "x" passes through channel H then it is converted to "y". The latter is shown in Equation (5) where n(f) is the noise and H(f) is the channel response [23].

$$y(f) = H(f)x(f) + n(f) \tag{5}$$

A channel's response includes Path Loss, Shadowing and Multipath effects. The Path Loss the dispersion of energy which spreads radially in space (starting from the point of emission) and in real conditions has been found that the signal is inversely proportional to the third or in many cases with the fourth power of the distance. Also, Shadowing is caused by the various obstacles which are located in the signal's path such as trees, buildings, etc. A consequence of these obstacles is the signal loss because the signal can be partially absorbed, scattered or even reflected and diffracted. Finally, the multipath delay spread is related to the fact that a receiver "sees" each incoming signal as a combination of various multipath signals. The signals (coming from an originally transmitted signal - from the transmitter's antenna) arrive at the receiver from multiple paths. These signals have different values of amplitude and phase due to various obstacles and different paths. The obstacles create scattering effects and consequently different arrival times of the scattered versions (of the original signal) at the receiver causing a delay in signal's reception. Its maximum value is named "maximum delay spread" and is the limit where the received signal becomes negligible. A larger value of this delay characterizes a more dispersive and frequency selective channel with a smaller coherence bandwidth. The coherence bandwidth (where the channel characteristics do not change) is inversely proportional to the delay spread. If the delay spread is large enough, then the symbol energy will spread into near symbols and it will lead to ISI (intersymbol interference). The coherence time is the time for which the channel is considered to be unchanged. Also, the frequency selective fading (wideband channel) causes undesirable and severe effects in communication systems with the wrong modulation techniques. [23,28,29].

When two systems move relatively to each other, then the frequency which is received is different compared to that of the primary transmitted signal. When the two systems move towards each other this will lead to the increase in the frequency and in the opposite occasion the frequency decreases. This phenomenon is called Doppler Effect and this is described in Equation (6).

$$\Delta f = \pm \frac{v_{RS}}{c} f_S .\tag{6}$$

where $\Delta f$ is the frequency change, $f_S$ the source frequency, c the speed of light and $v_{RS}$ the receiver's velocity relative to the source [30].

The Doppler spectrum is created by the presence and movement of various objects between the systems of transmitter and receiver inside the channel. The power distribution of the selected channel is inseparably connected to the Doppler spectrum (for one frequency). The bandwidth ranges from f-$f_D$ to f+$f_D$, where $f_D$ is called maximum Doppler spread (or Doppler spread). The Doppler spread calculates the spectral expansion due to channel's time rate of change. Also, this kind of spread is inversely proportional to the coherence time as shown by the Equation (7) [23,31].

$$\textit{Coherence time} \approx 1 / \textit{Doppler spread} \tag{7}$$

Mainly, the coherence time can be considered as the time duration that the characteristics of the channel do not change. The coherence time can be calculated as follows: If the spatial correlation length ($x_{CORR}$) equals to $\lambda$, c is the velocity of light, $f_O$ is the working frequency and u is the speed of the receiver, then the correlation time (often called coherence time) appears in Equation (8) where $V_{MAX} = (uf_O)/c$. If the term $V_{MAX}$ is replaced by $(uf_O)/c$ in Equation (8) then Equation (9) is produced [16].

$$t_{CORR} = x_{CORR} / u = 1 / V_{MAX} \tag{8}$$

$$t_{CORR} = c/(f_O u) \approx \frac{3\cdot 10^8 \cdot 3600\, Hz}{f_O} \frac{(km/h)}{u} \sec$$
$$t_{CORR} = \frac{1080\, MHz}{f_O} \frac{(km/h)}{u} \sec \qquad (9)$$

The term "fading" refers each time to an appropriate channel model (i.e. Rayleigh fading) and to its characteristics. The respective channel model (mathematical relationship) presents the variation of the signal while passing through the channel and expresses mathematically the particular type of statistical distribution of the signal. When line-of-sight route is dominant over the indirect signal paths, then this distribution is called Ricean. The inverse condition where the dominant component is the non-line-of-sight is called Rayleigh fading [28]. This type of fading is characterized by specific amplitude fading. The amplitude has a Rayleigh probability distribution. The Rayleigh distribution consists of two independent Gaussian components with a mean value of zero. The phase ranges from 0 to $2\pi$ rads and has a uniform distribution. If the two Gaussian distributions have equal standard deviations σ then, the Rayleigh fading amplitude is expressed by Equation (10). The Pdf (Probability density function) of the Ricean fading amplitude is shown in Equation (11). When the line-of-sight is dominant, then this condition corresponds to Ricean distribution as mentioned before. In this case, neither phase neither deviation have zero mean values and the channel is consisted of dominant paths (which contain most of the channel's energy) with multipath fading processes (without line-of-sight). In Equation (11), the parameter $\rho^2$ is the received power of non-fading components and $I_0$ is the modified Bessel function of the first kind. The Rice distribution may include AWGN and Rayleigh distribution introducing the K-factor, which is essentially the ratio of line-of-sight to non-line-of-sight, $K=\rho^2/2\sigma^2$ (Equation 12) [32,33].

$$f_{RAYLEIGH}(a) = \begin{pmatrix} \frac{a}{\sigma^2}\exp\left(-\frac{a^2}{2\sigma^2}\right) & \text{for } a \geq 0 \\ 0 & \text{for } a < 0 \end{pmatrix} \qquad (10)$$

$$f_{RICE}(a) = \begin{pmatrix} \frac{a}{\sigma^2}\exp\left(-\frac{a^2+\rho^2}{2\sigma^2}\right) I_O\left(\frac{a\rho}{\sigma^2}\right) & \text{for } a \geq 0 \\ 0 & \text{for } a < 0 \end{pmatrix} \qquad (11)$$

$$K = \frac{\rho^2}{2\sigma^2}, \quad \begin{pmatrix} K = 0, & Rayleigh \\ K \to \infty, & AWGN \\ K > 0 & \end{pmatrix} \qquad (12)$$

The K-factor affects the system's performance and should always be taken into consideration. From the Equation (12) is clear that when K has a value near zero, then this case corresponds to a Rayleigh distribution. On the contrary, when K is large enough to be considered as infinity then the whole energy will be included in only one path (LOS) and consequently it can be assumed that the model is an AWGN channel. Also, if K has a positive value (but not infinite) then it can be assumed having a Rice distribution (Equation 12) [10,32,34].

## 5.  Results

The simulated platform of noise AWGN, Rice, Rayleigh and phase noise consisted of an OFDM system (IFFT output = 2048 carriers) with a Turbo coding technique [9,11]. The selected coding scheme contained convolutional encoders, each one with a code rate of 1/2. The final code rate was equal to 1/4. Moreover, the fundamental encoders of the system had a constraint length of 3, memory of 2, generator polynomials $7_8$ and $5_8$ (octal form) and a feedback loop of 7. Also, it must be mentioned that the constraint length equals to the memory plus the input. As for the Turbo decoder, this stage included various APP decoders. These decoders were adjusted to work with Max* option which corresponded to log-MAP algorithm. This algorithm is shown in Equation (13).

$$V_{MAX} \max{}^*(X,Y) = \max(X,Y) + \log(1 + e^{-|Y-X|}) \qquad (13)$$

The simulated system is shown in Fig. 5 which consists of the transmitter, the receiver and the channel. The receiver's part includes all the grey blocks. Also, the two blocks named as "PAD, FT, IFFT, CP" and "REMOVE CP, FFT, REMOVE PAD" are presented in details in Fig. 6. The mentioned stages include the functions of zero padding, transferring the padding in the center of the subcarriers (Frame transformation before IFFT – FT, Frame Transformation), the Inverse Fourier Transform, Cyclic Prefix and then the opposite procedures.

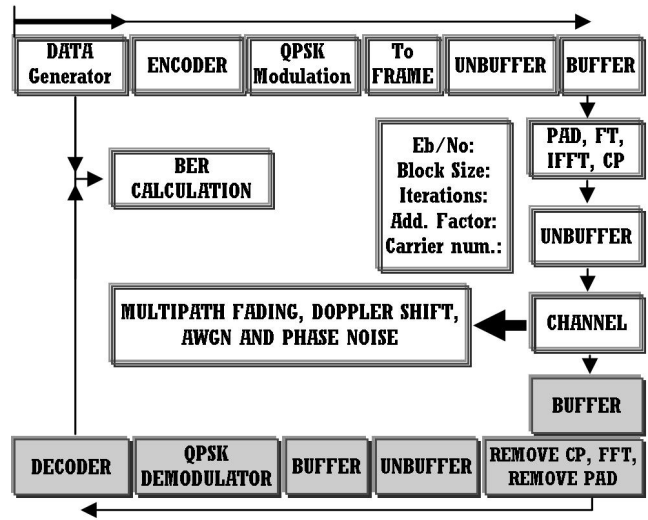

Fig. 5. Design of the simulated system

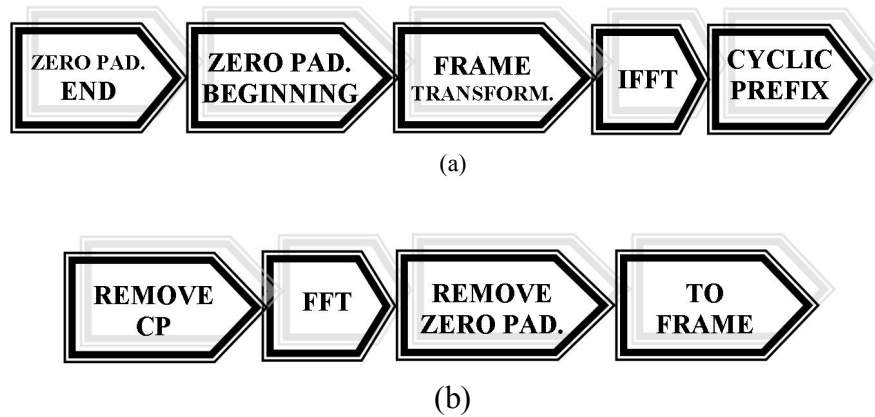

(a)

(b)

Fig. 6. Stages named (a) "PAD, FT, IFFT, CP" and (b) "REMOVE CP, FFT, REMOVE PAD"

Moreover, the channel model is constituted by Multipath fading, Doppler shift, AWGN and phase noise. The channel model is shown in Fig. 7 where there is an addition of various blocks in order to include the function of analyzing the original transmitted OFDM signal to the components of amplitude and angle. At the stage output of multipath fading, the previous signals are combined again in order to form a complex signal (OFDM), assuming that a perfect angle estimation exists. All simulations were conducted without synchronizing transmitter with the receiver because all resulting delays were taken into account for the proper functioning of the system.

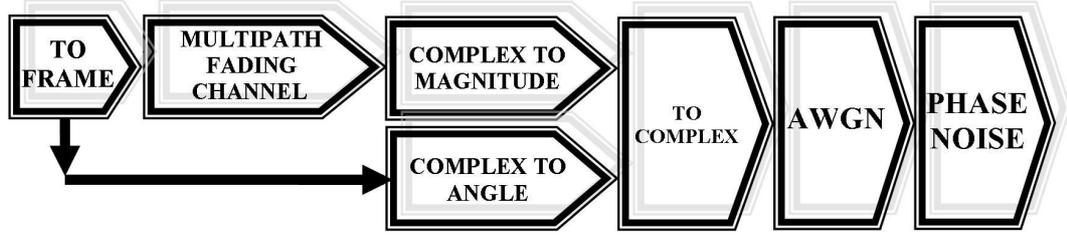

**Fig. 7. Noise model**

The amount of the inserted noise to the signal was adjusted each time by the signal-to-noise ratio per bit ($E_b/N_o$). Also, the bits per symbol are directly associated with the type of modulation. The utilized modulation type was QPSK (2 bits per symbol) with Gray constellation ordering. The $R_m$ term refers to the number of bits which constitute a symbol ($R_m=\log_2 M$, where M is the number of symbol states). The coding system had a code rate of 1/4 ($R_c$) and the noise variance was equal to No/2 (No is the spectral density). The total noise quantity added to a signal in an AWGN channel is shown in Equation (14).

$$\sigma^2 = \frac{E_S}{2\,R_M R_C\,(E_b/N_0)} = \frac{E_S}{2\,(log_2 M)\,R_C\,(E_b/N_0)} \qquad (14)$$

Equation (14) for the particular case of our PCCC system [9,11] was equal to $E_s / [4/3(E_b/N_o)]$. Also, the phase noise was generated from almost 0 Hz to $\pm F_s/2$, where $F_s$ was the sampling frequency. The produced phase noise was a type of 1/f noise applied to the total range of frequencies [35]. The phase noise level was equal to -50dB$_c$/Hz at a frequency offset of 100 Hz. The fading processes which were inserted in the scenario included Rayleigh channel shift of 30 and 40 degrees. Moreover, multipath Ricean fading with Rayleigh fading processes were simulated. This kind of simulations based on the model of ITU Pedestrian-A at 3km/h. The relative delays and powers of the simulated multipath intensity profile are presented in **Table 1** [14,23,36]. Furthermore, the maximum utilized Doppler spread was equal to 5.55 Hz [29] which means that the channel was changing 5.55 times per second. This value can be calculated from Equation (7) and Equation (9) for velocity of 3 km/h (ITU PA3) and working frequency of 2 GHz (in Equation 15).

$$Doppler\ spread = \frac{f_O}{1080\,MHz}\frac{u}{(km/h)}Hz$$
$$Doppler\ spread = \frac{2000\,MHz}{1080\,MHz}\frac{3\,km/h}{(km/h)}Hz \quad (15)$$

$$S(f) = 1/2f_d, \quad |f| \leq f_d \quad (16)$$

Flat fading was the type of the used Doppler spectrum in the proposed simulation environment. When a signal propagates through this kind of channel it sustains its spectral characteristics. Nevertheless, the power of the signal which reaches the receiver continuously varies. This fluctuation is caused by the multipath effects. So, a flat fading channel is also known as amplitude varying channel. This channel has a constant gain and its phase response is linear. The requirement for all the previous is that the transmitted signal's bandwidth is smaller than the channel's bandwidth [30]. The normalized flat Doppler power spectrum is shown in Equation (16) where $f_d$ is the maximum Doppler frequency [37]. It must be mentioned that the channel's effective gain for all paths was equal to 0 dB, and K factor was simulated for the values of K=1 and K=2. The fact of sustaining constant the K factor was that this could be applied to moving pedestrians with low speed. Furthermore, the two values of the K-factor were selected for being the worst integer values of so-called Low and High K-factor as reported in the literature [38]. Moreover, various values ranging from K=7 to K=0 had been examined by the literature [39] concluding in finding the appropriate K values which were the values K=1 and K=2. Also, a standalone Rayleigh channel (K=0) was simulated and produced the worst values of phase shifts. Major simulations settings are presented in **Table 2**.

**Table 1. Profile of ITU Pedestrian-A Channel**

| Relative Delay (ns) | 0 | 110 | 190 | 410 |
|---|---|---|---|---|
| Relative Power (dB) | 0.0 | -9.7 | -19.2 | -22.8 |

Table 2. Simulations settings

| Block name | Settings | Block name | Settings |
|---|---|---|---|
| Binary Generator | 50% probability of zero | Padding | 25% (End and Beginning) |
| IFFT output | Division by FFT length | Cyclic Prefix | 25% |
| QPSK Modulator and Demodulator | Phase offset of pi/4, Gray constellation ordering and Hard Decision type | Zero Order Hold (Fig. 3) and Gain (Fig. 3) | ZOH = ITER / (ADD. FACTOR) Gain = 2/VAR |
| Buffer (Fig. 5) | 0.75 CAR | Buffer 2 (Fig. 5) | 1.25 CAR |
| Buffer 3 (Fig. 5) | 2 LEN | Gain in Grey Color (Fig. 3) | 1/2 |

CAR is the number of Carriers, ITER is the simulated number of Iterations, LEN refers to the Block size, VAR is the Noise Variance and ADD. FACTOR ensures the same amount of data will be transmitted for different block sizes.

The Figs 8-11 show the performance of the Turbo Coded OFDM system which was simulated for various noise combinations. In Fig.8 are presented the simulation results for the innovative PCCC coding scheme [11] under the presence of AWGN, AWGN with phase noise and AWGN with Rayleigh channel shift of 30pi/180 rads block size. In all cases, block size is 512 and the number of iterations varies from 1 to 5. In the first two cases it is obvious that less than 0.6dB ($E_b/N_o$) is needed in order to accomplish a BER less than $10^{-6}$ which yields an improvement of ten thousand compared to the results of a conventional encoding scheme. In the third case, the previous BER can be sustained with almost 4dB increase in the $E_b/N_o$. In Fig.9, simulation results are presented for AWGN with Rayleigh channel shift of 40pi,/180 rads, AWGN with Rayleigh fading based on ITU PA3 and AWGN with Ricean fading and Rayleigh processes based on ITU PA3 (K=1). Less than 12.5 dB ($E_b/N_o$) were needed in order to be achieved a BER of $10^{-3}$. This refers to the case of Rayleigh channel shift of 40pi/180 rads. All other cases required above 7 dB ($E_b/N_o$) for exhibiting the previous BER. The same conclusion can be reached for the case results shown in Fig.s 10a to 10c. In other words we need at least 7 dB ($E_b/N_o$) to accomplish a BER of $10^{-3}$. Finally, in Fig.11 are shown all simulated noise types for 512 block sizes and 5 iterations. As shown, the case of an AWGN with Rayleigh fading (PA3) is identical to the case of an AWGN with PN and Ricean fading with Rayleigh processes (PA3 and K=1).

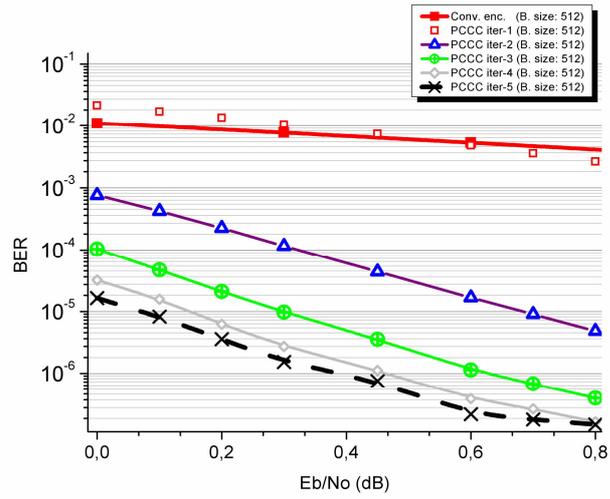

(a)

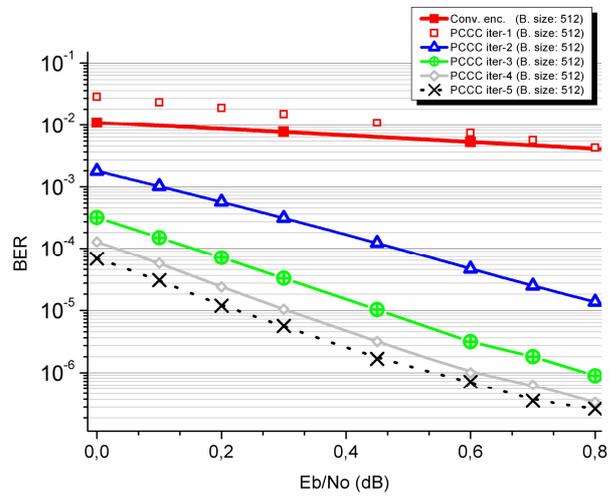

(b)

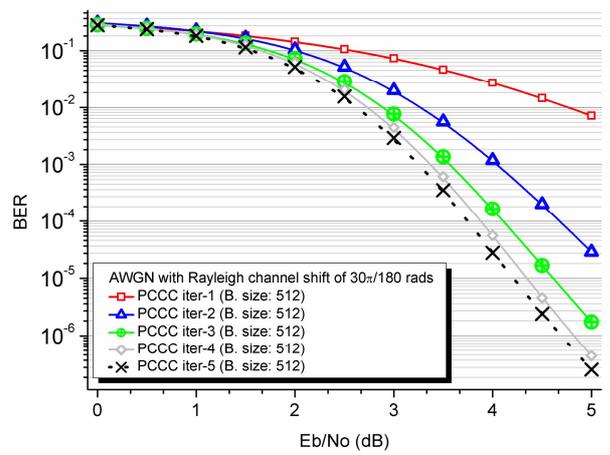

(c)

**Fig. 8. (a) AWGN, (b) AWGN with Phase noise and (c) AWGN with Rayleigh channel shift of 30pi/180 rads.**

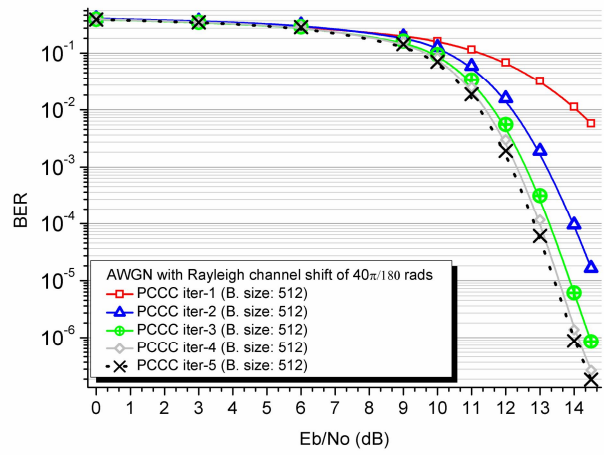

(a)

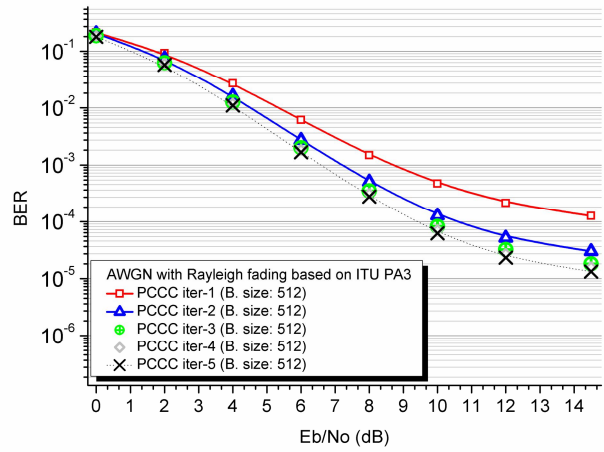

(b)

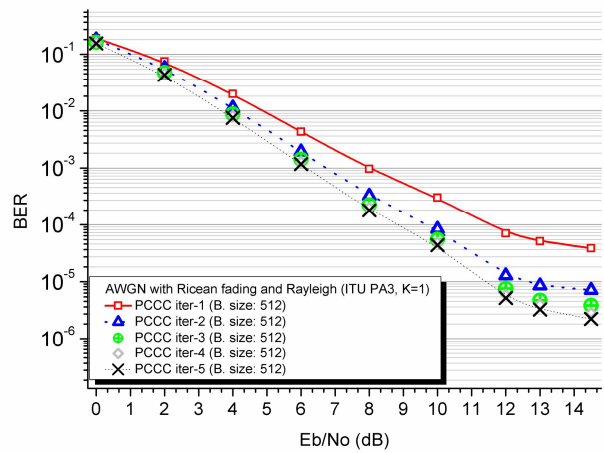

(c)

Fig. 9. (a) AWGN with Rayleigh channel shift of 40pi,/180 rads (b) AWGN with Rayleigh fading based on ITU PA3 and (c) AWGN with Ricean fading and Ralyleigh processes based on ITU PA3 (K=1).

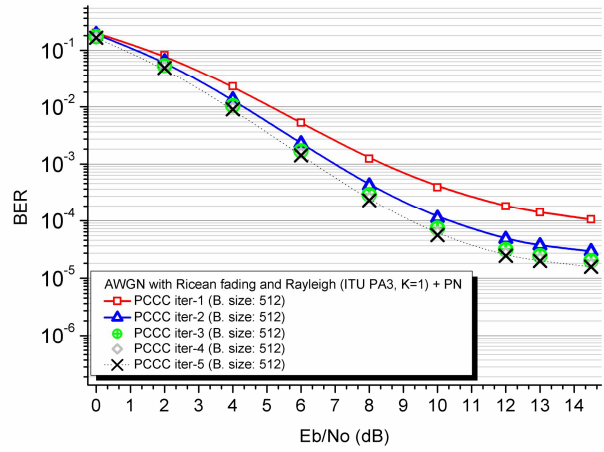

(a)

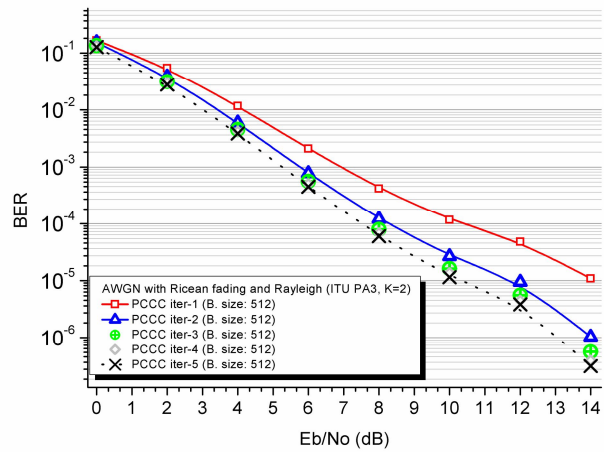

(b)

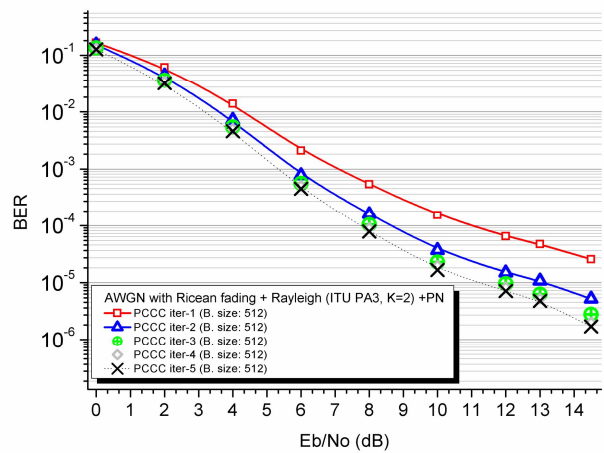

(c)

**Fig. 10. (a) AWGN with Ricean fading and Ralyleigh processes based on ITU PA3 (K=1) and with PN, (b) AWGN with Ricean fading and Ralyleigh processes based on ITU PA3 (K=2) and (c) AWGN with Ricean fading and Ralyleigh processes based on ITU PA3 (K=2) and with PN.**

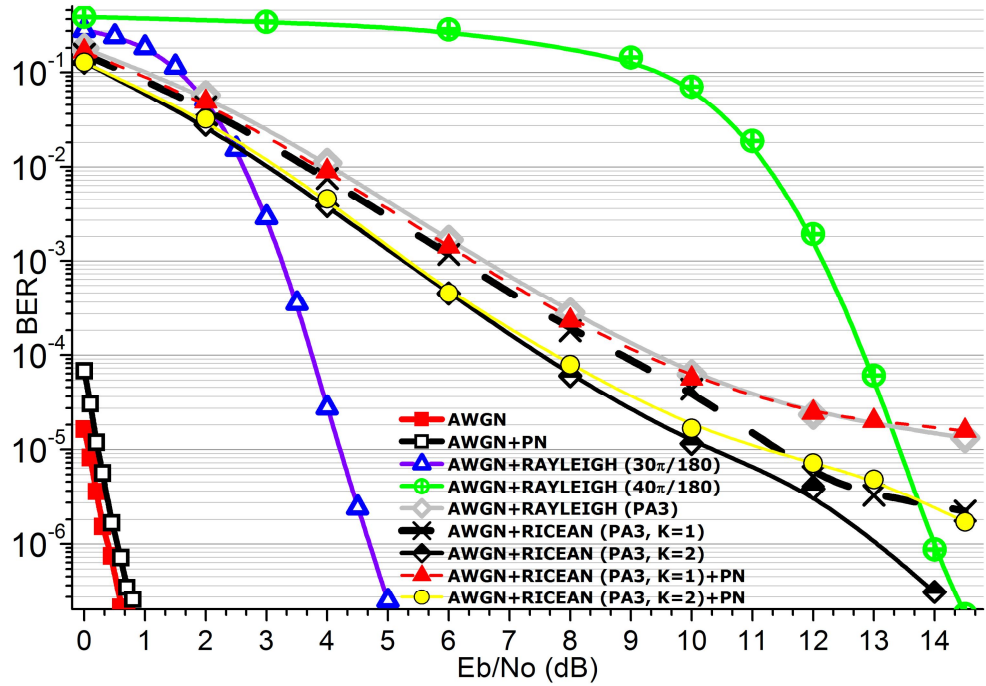

Fig. 11. All simulated type of noises for 5 iterations and 512 block size.

## 6. Conclusions and future work

In this paper, an innovative Turbo Coded OFDM system [9,11] was evaluated under several realistic noise scenarios. Simulation results show that less than 12.5 dB ($E_b/N_o$) were needed in order to achieve a BER of $10^{-3}$ which refers to the worst case of Rayleigh channel shift of 40pi/180 rads. All other scenarios required at least 7 dB ($E_b/N_o$) for exhibiting the previous BER.

This system is intended to be a part of systems with incorporated technologies such as UWB, MIMO and tunable antennas. The full system's implementation on a DSP platform along with channel estimation techniques similar to [40] will include noise effects under real conditions.

## Acknowledgment

This research has been co-financed by the European Union (European Social Fund – ESF) and Greek national funds through the Operational Program "Education and Lifelong Learning" of the National Strategic Reference Framework (NSRF) - Research Funding Program: Heracleitus II. Investing in knowledge society through the European Social Fund.